 \documentstyle[12pt]{article}
\newlength{\abstractwidth}
\renewcommand{\thefootnote}{\fnsymbol{footnote}}
\renewcommand{\thanks}[1]{\footnote{#1}} 
\newcommand{\starttext}{
\setcounter{footnote}{0}
\renewcommand{\thefootnote}{\arabic{footnote}}}

\newcommand{\be}{\begin{equation}}
\newcommand{\bea}{\begin{eqnarray}}
\newcommand{\eea}{\end{eqnarray}}
\newcommand{\beq}{\begin{equation}}
\newcommand{\ee}{\end{equation}}
\newcommand{\eeq}{\end{equation}}
\newcommand{\N}{{\cal N}}
\newcommand{\<}{\langle}

\renewcommand{\a}{\alpha}
\renewcommand{\b}{\beta}
\newcommand{\m}{\mu}
\newcommand{\n}{\nu}
\newcommand{\mf}[1]{m_\phi^2(k_{#1})}
\newcommand{\mt}[1]{m_t^2(k_{#1})}
\newcommand{\half}{{1\over 2}}
\newcommand{\cabc}{\<C^{k_1}C^{k_2}C^{k_3}\>}

\renewcommand{\>}{\rangle}
\def\ba{\begin{eqnarray}}
\def\ea{\end{eqnarray}}

\def\N{{\cal N}}
\def\O{{\cal O}}

\def\tr{{\rm tr}}
\def\12{{1 \over 2}}
\def\32{{3 \over 2}}
\def\72{{7 \over 2}}
\def\92{{9 \over 2}}

\def\AdS{{\rm AdS}}

\begin{document}
\baselineskip=18pt

\vfill\eject

\begin{titlepage}
\bigskip
\hskip 3.7in\vbox{\baselineskip12pt
\hbox{MIT--CTP--2893}
\hbox{UCLA/99/TEP/27}
\hbox{hep-th/9908160}}
\bigskip\bigskip\bigskip\bigskip

\centerline{\Large \bf Extremal correlators in the AdS/CFT
correspondence  }
\bigskip\bigskip
\bigskip\bigskip

\centerline{ Eric D'Hoker$^{a}$, Daniel Z. Freedman$^{b,c}$, 
Samir D. Mathur$^{b}$} 

\medskip
 
\centerline{Alec Matusis$^{b}$ and Leonardo Rastelli$^{b,}$\footnote[1]{\tt 
e-mails : dhoker@physics.ucla.edu, dzf@math.mit.edu, me@ctpdown.mit.edu, 
alec\_m@ctp.mit.edu, rastelli@ctp.mit.edu} }
\bigskip
\bigskip
\centerline{$^a$ \it Department of Physics}
\centerline{ \it University of California, Los Angeles, CA 90095}
\centerline{\it and  Institute for Theoretical Physics}
\centerline{ \it University of California, Santa Barbara, CA 93106}  
\bigskip
\centerline{$^b$ \it Center for Theoretical Physics}
\centerline{$^c$ \it Department of Mathematics}
\centerline{ \it Massachusetts Institute of Technology}
\centerline{\it Cambridge, {\rm MA} 02139}
\bigskip\bigskip

\begin{abstract}

\medskip

The non-renormalization of the 3-point functions 
$\<\tr  X^{k_1} \ \tr  X^{k_2} \ \tr X^{k_3}\>$ 
of chiral primary operators in
$\N=4$ super-Yang-Mills theory is one of the most striking facts to emerge from
the AdS/CFT correspondence. A two--fold puzzle appears in the extremal
case, e.g. $k_1=k_2+k_3$. First, the supergravity calculation involves
analytic continuation in the $k_i$ variables to define the product of a 
vanishing
bulk coupling and an infinite integral over AdS. Second, extremal
correlators are uniquely sensitive to mixing of the single--trace operators 
$\tr X^k$ with protected multi-trace operators in the same representation
of $SU(4)$. We show that the calculation of extremal correlators from
supergravity is subject to the same subtley of regularization known for
2-point functions, and we present a careful method which justifies the 
analytic continuation and shows
that supergravity fields couple to single traces without admixture. We
also study extremal $n$-point functions of chiral primary operators, and argue
that type IIB supergravity requires that their space-time form is a product of
$n-1$ two--point  functions (as in the free--field approximation) multiplied 
by a
non--renormalized  coefficient. This non--renormalization property of extremal
$n$--point  functions is a new prediction of the AdS/CFT correspondence.
As a by-product of this work we obtain the cubic couplings $t\phi\phi$ and
$s\phi\phi$ of fields in the dilaton and 5-sphere graviton towers of 
type IIB supergravity on $AdS_5 \times S_5$.

\end{abstract}

\end{titlepage}

\starttext
\baselineskip=16pt
\setcounter{equation}{0}
\setcounter{footnote}{0}

\section{Introduction}

The study of correlation functions of operators in the boundary theory
is one useful way to explore the AdS/CFT correspondence 
\cite{maldacena, polyakov, witten}  (see 
\cite{review} for a comprehensive review). When applied to Type
IIB supergravity on $AdS_5 \times S^5$ this study has uncovered
several previously unknown properties of
the $d=4$, $\N=4$, super--Yang--Mills (SYM) theory with gauge group $SU(N)$.
In the important calculation of \cite{seiberg}, 
3--point functions of  the $\N=4$ chiral primary
operators $\tr X^k \equiv \tr \{ X^{i_1}(x) X^{i_2}(x) \cdots X^{i_k}(x) \}$
(in the symmetric traceless representation of $SU(4)$ with Dynkin label $(0,k,0)$)
were computed at strong coupling using supergravity,
 and found to exactly agree at large $N$ with
the free field approximation to the SYM theory.

Upon closer examination of the computation of \cite{seiberg}, one
finds a curious puzzle in the case
of `extremal' 3--point functions, namely correlators 
$\<\tr  X^{k_1}\tr  X^{k_2}\tr X^{k_3}\>$ in which the conformal
dimension of one of the operators is precisely equal
to the sum of the other two dimensions, e.g. $k_1=k_2+k_3$.
(These values of $k_i$ are  `extremal' in the sense that for 
$k_1 >k_2+k_3$ the 3--point function vanishes by $SU(4)$
selection rules.)
After dimensional reduction of the field equations of the 
Type IIB supergravity,
scalar fields $s'_k$ were defined and found to have
a simple cubic interaction of the form
${\cal G}(k_1, k_2, k_3) s'_{k_1} s'_{k_2} s'_{k_3}$.
In the extremal case $k_1 =k_2 +k_3$ the coupling
${\cal G}(k_2+k_3, k_2, k_3)$ vanishes but the
integral over $AdS_5$ needed to compute the
3--point functions diverges \cite{april}.

One way around the problem (which was implicitly
used in \cite{seiberg} and \cite{texas} and that has been 
recently emphasized in \cite{liutseytlin3})
is to analytically continue in the conformal
dimensions $k_1$, $k_2$, $k_3$. As $k_1 \rightarrow k_2 + k_3$ the AdS integral
computed in \cite{april} diverges as
$1/(k_1-k_2-k_3)$. In the same limit the supergravity
coupling goes to zero as $k_1-k_2-k_3$. Thus by
analytic continuation one obtains a finite result
for the extremal 3--point functions. This is the answer
quoted in \cite{seiberg} which matches the free field result.
Although analytic continuation may be regarded as a regularization 
procedure, it  lacks rigourous justification since $SU(4)$ symmetry
requires integer dimensions and thus integer $k_i$. Additionally, the singularity
of the integral over $AdS_5$ arises from the region where the
bulk interaction point approaches the boundary. In similar situations
encountered previously \cite{polyakov,april} a
regularization procedure using a space-time cutoff inside the boundary
was used and found to give satisfactory results. In this case, however,
the extremal correlators would vanish for any finite value of the cutoff
and also in the limit as the cutoff is removed. Thus the puzzle is to
find a method to compute the extremal correlators
from first principles, working 
at exact extremality with
definite integer values of $k_2$, $k_3$, $k_1=k_2+k_3$, and we propose
such a method in this paper.

A further motivation for considering extremal correlators 
arises from field theory. It is customarily
believed that the supergravity fields $s'_k$ are dual to
the single trace operators $\tr X^k$. Indeed both $s'_k$ and
$\tr X^k$ belong to the same short representation
of the governing superalgebra $SU(2,2|4)$ and have protected scale
dimension $\Delta=k$.
However, it has been recognized \cite{ferrara,bianchi,skibanew}, 
but not widely considered so far, that certain 
multi--trace operators are also BPS operators with protected dimensions.
In addition to the single trace chiral primary
$\tr \{ X^{i_1}(x) X^{i_2}(x) \cdots X^{i_k}(x) \}$ 
one may  consider, for example,
double trace BPS operators of the form
$\tr \{ X^{i_1}(x) X^{i_2}(x) 
\cdots X^{i_{k-l}}(x) \} \tr \{ X^{j_1}(x) X^{j_2}(x) \cdots X^{j_l}(x) \}$, 
in the same $(0,k,0)$ 
$SU(4)$ representation. Projection into this representation is obtained
by the same process of total symmetrization in the indices $i_m$ and
removal of traces that is used for $\tr X^k$ itself. More generally, one
can consider higher order multi--trace operators, schematically denoted
by $\tr X^{k_1}\tr X^{k_2}\cdots\tr X^{k_m}$, projected into the representation
$(0,k=k_1+k_2+\cdots k_m,0)$. All of these operators transform in the same
representation of $SU(2,2|4)$ as the single trace $\tr X^k$,  so we would 
expect these operators to mix, and, indeed, mixing is required by the general
structure of operator product expansions in the ${\rm SYM}$ theory.
(The relevance of multi--trace operators for the $AdS_5/CFT_4$
correspondence on the Coulomb branch has been noted in \cite{trivedi}
and \cite{klebwit}.)

One is thus naturally led to speculate that supergravity fields
couple to linear combinations of single and multiple
trace BPS operators. In the free field approximation, one can easily
see that operator mixing occurs, although generically
suppressed by powers of $1/N$.
Thus one
might get the impression that the issue of whether supergravity fields couple
to single trace operators or to admixtures with protected multi-traces is
a non-leading effect of secondary concern in the AdS/CFT correspondence.
However, extremal correlators are exceptional: the contribution
of multi--trace operators to extremal 3--point functions is enhanced and
of the same order in $N$ as single traces. It is quite curious that multi-trace
admixtures are important in exactly the same situation where there is an 
ambiguity in the calculation of correlation functions.
Thus in order to decide whether
multi--trace admixtures are  present
in the AdS/CFT correspondence,
it is 
crucial to have a reliable scheme for computing
extremal correlators in supergravity.

In this paper, we will address these issues through a concrete
supergravity calculation. We will compute the 3--point
functions 
$\< \O_t^{k_1} \O_\phi^{k_2} \O_\phi^{k_3} \>$, where
$\O_\phi^{k}$ and $\O_t^{k}$ are the SYM operators that couple 
respectively to Kaluza--Klein
modes of the dilaton $\phi$ and of  the supergravity
scalar  field $t$  (a linear combination of the 4--form and of the trace of 
the graviton with indices on the $S^5$). One motivation
 for considering this example (which has also been 
recently studied in \cite{liutseytlin3})  
is that the details of the dimensional reduction are somewhat simpler
than in the chiral primary computation of \cite{seiberg}.
In particular, no subtlety related to the self--duality
condition of the 4--form arises, and we will be able to work
always at the level of an explicit action. Since the 3-point functions
of primary and descendent operators are related by supersymmetry
transformations, it is quite clear that our method can be applied
to primary correlators also.

It should be noted that two sets of scalar fields were considered in
\cite{seiberg}. The dimensional reduction procedure initially leads
to equations of motion for fields $s_k$ with both
derivative and non--derivative interactions.
A nonlinear transformation to the $s'_k$ fields then eliminates the
derivative couplings. In the descendent sector we work with fields
which are analogous to the $s_k$ and have both derivative and non--derivative
couplings. The interaction Lagrangian is then manipulated by partial
integration and use of the linear equations of motion. The value of the 
on--shell action (as a functional of boundary data)
is expressed as the sum of  non--derivative cubic bulk coupling plus
certain cubic couplings which are total derivatives or boundary interactions.
We then compute 3-point correlation functions using the two methods of
analytic continuation and space--time cutoff.

We find that in the non--extremal cases $k_1< k_2+k_3$ the contribution
of the boundary interactions vanishes as the cutoff is removed, and the net
contribution of the bulk vertex is completely unambiguous. At
extremality, however, the bulk coupling constants vanish as in the 
calculation of \cite{seiberg} in the chiral--primary sector. 
So the cutoff
method gives a vanishing bulk contribution, but one of the boundary vertices  
gives a non--vanishing result. This result is then compared with the
limit  $k_1 \rightarrow k_2+k_3$ of the analytically continued non--extremal
correlator, and we find exact agreement.


Our results provide a justification, at least in this example,
 for the analytic continuation
procedure used in \cite{seiberg} and advocated in \cite{liutseytlin3}.
We expect a very similar situation to occur in the chiral primary
computation of \cite{seiberg}. Namely, the 
extremal 3--point
functions could  in principle be
obtained from a boundary interaction
of the original fields $s_k$ 
that directly arise from the dimensional reduction.
In the change of variables to the new fields $s'_k$
(equ. (3.36) of \cite{seiberg}), 
these boundary terms are removed and the naive coupling appears to be zero
in the extremal case. 
However, we expect analytic continuation to extremality 
to  give the same answer as a calculation at exact extremality with
boundary interactions.

Our analysis thus indicates that the fields $s_k$ couple to single
trace operators only, although admixtures of multi--trace operators
might have been expected. 
If the new procedure had led to values of extremal correlators
different from those obtained from analytic continuation then 
one would have to add multi-trace admixtures to explain the
difference. Since this does not happen, we must
conclude that the customary map to single trace
operators is correct, or that if present double--traces are 
suppressed (like $1/N^2$).


We also consider extremal $n$--point functions of chiral primary
operators. Analysis of the supergravity
interactions allows us to prove (under reasonable
assumption about the supergravity Lagrangian)
that the functional form (that is, the dependence on space--time coordinates)
of these correlators in the strong coupling limit
is the same
as in the free field approximation. Arguments based on the operator
product expansion can then be used to show that the coefficient
of this functional form satisfies a non--renormalization theorem and
is independent of the coupling $\lambda = g^2N$.

A curious subtlety emerges from our calculation of extremal
3-point functions. The initially derived interaction Lagrangian
for the bulk fields is the sum of two bulk vertices, one with
two derivatives and one with four derivatives. Each vertex
contributes to an extremal 3-point function via a bulk integral
which diverges logaritmically as the bulk point $z_\mu=(z_0,z_i)$
approaces the boundary at $z_o=0$. The divergence cancels in the
sum of the two integrals, but regularization by a simple cutoff
at $w_0=\epsilon$ does not give the same result for the correlator
as the method described above in which Dirichlet boundary conditions
were imposed. Thus extremal 3-point functions are a new example of
the subtlety in the regularization of 2-point functions discussed
in \cite{witten,april,klebwit}. Formally 2-point functions,
eg $< O_\phi^{k_1}  O_\phi^{k_2}>$, are also extremal, since $k_1=k_2$
is required by conformal symmetry.

The paper is organized as follows. In Section 2 we present some results,
part new and part review, concerning the mixing of single and multi-trace
operators in the ${\rm SYM}$ field theory. Section 3 contains our treatment of
3-point correlators in the $t-\phi-\phi$ descendent sector of type
$IIB$ supegravity, and Section 4 describes the extension of our ideas
to extremal $n$--point functions.

\setcounter{equation}{0}
\setcounter{footnote}{0}

\section{Multi--trace BPS operators and Mixing}

We denote normalized single trace chiral primary operators by
\be \label{Ok}
\O _k (x) = \frac{1}{N^{k/2}}\tr \{ X^{i_1}(x) X^{i_2}(x) \cdots X^{i_k}(x) \}
\ee
in which it is understood that the tensor is totally symmetric and traceless
in the $SU(4)$  indices $i_\ell=1,\cdots ,6$.
The double-trace product of two such operators  
$\O_j (x) \O_{k-j}(x)$ transforms as the direct sum of the irreducible 
representations of $SU(4)$ in the tensor product $(0,j,0) \otimes (0,k-j,0)$.
The tensor product always contains, with unit multiplicity, the representation
with Dynkin label $(0,k,0)$, and the projection into this
representation is obtained by symmetrization and removal of traces. Similar
remarks apply to the order $m$ multi-trace operator 
$\O_{k_1}\O_{k_2}\cdots \O_{k_m}$, with $k=k_1+k_2+\cdots + k_m$, which also
contains the representation $(0,k,0)$ with unit multiplicity. We denote
these highest weight components of the direct product by
\be
[\O_k (x) \O_\ell(x) ]_{\rm max} 
\equiv [\O_k (x) \O_\ell (x)]\bigg | _{(0,k+\ell,0)}.
\ee
\be 
[\O_{k_1} (x) \cdots  \O_{k_m}(x) ]_{\rm max} 
\equiv [\O_{k_1} (x) \cdots \O_{k_m} (x)]\bigg | _{(0,k_1 + \cdots + k_m,0)}
\ee

There is a simple BPS argument that all of these operators 
transform in the unique short representation of $SU(2,2|4)$
whose lowest weight component is a scalar operator in the 
$(0,k,0)$ representation of $SU(4)$, of scale dimsnsion 
$\Delta =k$. Indeed, these operators must be lowest weight,
since they are totally symmetric in flavor indices, and 
there can be no appearance of descendents by using the field
equations of the interacting SYM theory.

A complete classification of unitary representations of $SU(2,2|4)$
was given in \cite{dobrev}. For the representations with $SU(4)$ quantum
numbers $(0,k,0)$ of the lowest dimension (scalar) component in the 
multiplet, two possible types appear : (1) there is a BPS multiplet 
with  dimension $\Delta = k$, (2) there is a continuous family
of representations with $\Delta \geq k+2$. The above operators
$[\O_{k_1} \cdots \O_{k_m}]$
all have the property that at vanishing SYM coupling, $\Delta =k$,
and thus are in representations of the type (1).
Assuming continuity of the dimension as a function of $g$,
we see that the operators have to be of type (1) for all
values of $g$, and are thus BPS. A somewhat different argument was given
in \cite{skibanew}.

Next, we show that mixing of single-- and double--trace operators is a rigorous
consequence of the operator product expansion in conformal field theory.
First we observe that the operator product expansion $\O_{k_2}(y)\O_{k_3}(z)$
contains the (non-leading) term
\be 
O_{k_2}(y)\O_{k_3}(z) \longrightarrow [\O_{k_2}(y) \O_{k_3}(y)]_{\rm max}
\ee
with unit coefficient as $z \rightarrow y$, since the operator
$[\O_{k_2}(y) \O_{k_3}(y)]_{\rm max}$ is actually {\it defined} by the
short distance limit. From the $z\rightarrow y$ limit of the extremal 3--point 
function $(k_1=k_2+k_3)$
\be
\<\O_{k_1}(x)\O_{k_2}(y)\O_{k_3}(z)\> = 
   {c\over (x-y)^{2k_2}(x-z)^{2k_3}}
\ee
one finds the exact result
\be
\<\O_{k_1}(x)[\O_{k_2}(y)\O_{k_3}(y)]_{\rm max}\>=
    {c\over (x-y)^{2k_1}}
\ee
which shows that the mixed two-point function is non-vanishing and has
the same non--renormalization properties as the extremal 3--point function.
 
\subsection{Large $N$ counting}
We now apply large $N$ counting arguments to obtain the order in $N$ of
the various correlation functions of interest in this paper. To do this
one can convert free field Feynman diagrams involving adjoint scalars
into fundamental planar `quark' diagrams in which each closed quark loop
gives a factor of $N$. Planar interaction diagrams then have the same
net power of $N$ if the 't Hooft coupling $\lambda=g^2N$ is fixed. Actually,
if the order $\lambda^2$ non--renormalization results of \cite{dhfskiba} for
single--trace correlators and the new results of \cite{skibanew} for 
correlators of protected multi--trace operators hold in higher order, then
free field diagrams tell the whole story. In the formulas below we give
the power of $N$ associated with the various correlators of interest in this
paper (other factors are usually omitted).

We normalize free field propagators as
\be
\< X^a (x) X^b(y) \> \sim \frac{\delta^{ab}}{(x-y)^2}
\ee
where $a,b$ are color indices. It then follows
that the operators $\O_k$ in (\ref{Ok}) have unit normalization
in the large $N$ limit
\be
\< \O_k(x) \O_k(y) \> \sim \frac{1}{(x-y)^{2k}} \,.
\ee
Large $N$ counting then tells us that mixed 2-point functions of single
trace operators and protected multi-traces are suppressed by further
powers of $N$, specifically (for $k=k_1+k_2$, etc.)
\bea
\<\O_{k}(x)[\O_{k_1}(y)\O_{k_2}(y)]_{\rm max}\> \sim {1\over N},
\nonumber \\
\<\O_{k}(x)[\O_{k_1}(y)\O_{k_2}(y)...\O_{k_m}(y)]_{\rm max}\>
\sim {1\over N^{m-1}}.
\eea
Thus there is operator mixing even at the free-field level in the SYM
theory. The matrix of 2-point functions is diagonalized by eigen-combinations
of operators with leading term $\O_K$ plus admixtures of 
$[\O_{k_1}\O_{k_2}...\O_{k_m}]_{\rm max}$ with coefficients of order
$1/N^{m-1}$. It is thus suggestive, although logically independent, that
supergravity fields $s_k$ or $s'_k$ couple to similar linear combinations
in the $AdS_5/CFT_4$ correspondence, and this hypothesis is explored in
the next section.

For 3-point correlators, large $N$ quark loop counting (and $SU(4)$ flavor
invariance) give the results
\bea
&&\< \O_{k_1} \O_{k_2} \O_{k_3} \> \left\{ \begin{array}{ll}  =0 
& {\rm if}\; k_1 > \;k_2 +k_3 
\\  \sim \frac{1}{N} & {\rm if} \;k_1 \leq k_2 +k_3 \end{array}\right.
\nonumber \\\nonumber \\
&&\< [\O_{l} \O_{k_1-l}]_{\rm max} \O_{k_2} \O_{k_3} \>
 \left\{  \begin{array}{ll} =  0  &  {\rm if} \; k_1 > k_2 +k_3 
\\  \sim 1  &  {\rm if}\;  k_1 = k_2 +k_3 \;{\rm and}\;
l=k_2 \;{\rm or} \;l =k_3  \\
\sim\frac{1}{N^2} &  {\rm if} \;k_1 <  k_2 +k_3 \;
{\rm or} \;k_1 =  k_2 +k_3 \, , \,  l \neq  k_2, k_3. \end{array}\right.
\eea
The last equation shows that the correlators
$\< \O_{k_1} \O_{k_2} \O_{k_3} \>$ and $\< \sum_l \frac{1}{N}\,[\O_{l} 
\O_{k_1-l}]_{\rm max} \O_{k_2} \O_{k_3} \>$
are  comparable in the extremal case $k_1 =k_2 +k_3$
(and the leading contribution in the second
one comes from $l=k_2$ or $l=k_3$), while for
$k_1 < k_2 + k_3$ the second correlator is $O(1/N^2)$
compared to the first. Thus extremal 3--point functions are special
in the sense that double--trace operators give an enhanced
contribution. This means that the agreement between supergravity and
free--field theory for non-extremal correlators does not test the
possible coupling to multi--trace operators, and a reliable method
of computation of extremal correlators is required.

A similar situation holds for (connected) extremal 4--point functions of 
single and multi--trace operators. For example
\be
\<\O_K \O_{k_1} \O_{k_2} \O_{k_3} \> \sim {1\over N^2}
\< [\O_{k'_1} \O_{k'_2} \O_{k'_3}]_{\rm max} \O_{k_1} \O_{k_2} \O_{k_3} \>
 \sim  \left\{ \begin{array}{ll} 1   &{\rm if}\;  k'_i = k_i \\
\frac{1}{N^3}    &{\rm if} \; k'_i \neq k_i 
\,.
\end{array}\right.
\ee

\setcounter{equation}{0}
\setcounter{footnote}{0}

\section{Supergravity computation of $ \< \O_{t,s}^{k_1}  \O_\phi^{k_2}  
\O_\phi^{k_3}  \>$  }

In this Section we describe the computation of the correlation functions
 $ \< \O_t^{k_1}  \O_\phi^{k_2}  \O_\phi^{k_3}  \>$ and
 $ \< \O_s^{k_1}  \O_\phi^{k_2}  \O_\phi^{k_3}  \>$, 
where   $\O_\phi^{k}$, $\O_t^{k}$, $\O_s^{k}$
are the SYM operators coupling respectively to the supergravity
scalar fields $\phi^k$ (KK modes of the dilaton),
$t^k$ and $s^k$ (scalar KK modes arising from the 4--form and the graviton
with indices on the sphere). 
In principle these correlators are related by supersymmetry
to the 3--point
functions of chiral primaries computed in \cite{seiberg}, although
the explicit relation is cumbersome to obtain.

Besides giving some new potentially useful explicit results for these
correlators, the computations presented here will allow us to settle some
questions of principle that arise for the `extremal' cases.
We will describe in some detail the calculation of  
$ \< \O_t^{k_1}  \O_\phi^{k_2}  \O_\phi^{k_3}  \>$, giving particular
 emphasis to the
subtleties that arise for the extremal  values $k_1=k_2+k_3$.
The computation of  $ \< \O_s^{k_1}  \O_\phi^{k_2}  \O_\phi^{k_3}  \>$ is
completely analogous and we will just quote the result.

\subsection{Set--up}

We closely follow \cite{vannieuv}. We adopt almost
uniformly the conventions introduced in \cite{seiberg},
which we recall here for the reader's convenience.
However, unlike \cite{vannieuv} and \cite{seiberg}
we will work with Euclidean signature.

We use latin indices $i,j,k,\dots$ for the whole 10--dimensional
manifold. Indices $\alpha,\beta,\gamma,\dots$ are $S^5$ indices,
while $\mu, \nu, \lambda, \dots$ are $AdS_5$ indices.
$G_{mn}$ indicates the metric and $g_{mn}$ its background value.
We set the $S^5$ and $AdS_5$ scales to 1, {\it i.e.}
we use units such that the Riemann tensor for the background solution
takes the form
\be
R_{\mu \lambda \nu \sigma} = -(g_{\mu \nu}g_{\lambda \sigma}-
g_{\mu \sigma}g_{\lambda \nu}), \quad
R_{\alpha \gamma \beta \delta} = (g_{\alpha \beta}g_{\gamma \delta}-
g_{\alpha \delta}g_{\gamma \beta})\,.
\ee

We set
\bea\label{fluct}
&&G_{mn} = g_{mn} + h_{mn},\\
&&h_{\a\b}=h_{(\a\b)}+{h_2 \over 5} ; \  \ g^{\a\b}h_{(\a\b)}=0, \\ 
&&h_{\m\n} = h'_{\m\n} -{h_2 \over 3} g_{\m\n},\  \
h'_{\m\n} = h'_{(\m\n)} + {h' \over 5} g_{\m\n};\  \ 
g^{\m\n}h'_{(\m\n)} = 0,\\
&&F=\bar{F} + \delta F,\  \
\delta F_{ijklm}=5D_{[i}a_{jklm]}.
\eea
Here, $\bar F$ is the background value of the $F$-field.
Following \cite{vannieuv}, we choose the gauge conditions
 $D^{\a}h_{\a\b}=D^{\a}h_{\m\a} = D^\a a_{\a\m_1m_2m_3m_4}=0$.
We refer to \cite{vannieuv} for a complete discussion
of gauge fixing and for the general expansion
of the fluctuation in harmonics of the sphere.
We will only need
\bea\label{sphdecomp}
{h'}_{\m\n}& =& \sum Y^{k} {h'}^k_{\m\n},\\
h_2 & =& \sum Y^{k}h_2^{k},
\\
a_{\a_1\a_2\a_3\a_4}& = & \sum D^{\a}Y^{k}\epsilon_{\a\a_1\a_2\a_3\a_4}b^{k},
\\
\phi&=&\sum Y^k \phi^k 
\eea

The modes $h_2^k$ and $b^k$ have coupled linear equations
of motion. The diagonal combinations are \cite{vannieuv} \cite{seiberg}
\bea\label{diag}
s^k& =&{1\over 20 (k+2)}[h_2^k-10(k+4)b^k], \label{defs}\\
t^k& =&{1\over 20 (k+2)}[h_2^k+10k{b^k}] \label{deft} 
\eea
which satisfy
\bea\label{emef}
D_{\m}D^{\m}s^k& =& k(k-4) s^k, \\
D_{\m}D^{\m}t^k& =& (k+4)(k+8) t^k.  
\eea

\subsection{Constraints}

The equations of motion that follow from supergravity 
yield some constraints 
between the modes defined above. 
It was shown in \cite{vannieuv}\cite{seiberg} that 
if we excite the field $h_2$ then the constraints
force us to also excite a 
certain amount of the field $h'_{\mu\nu}$. The traceless part and the trace 
part of this latter field are determined as follows
\bea {h'}^k_{(\m\n)}& =& D_{(\m}D_{\n)}
 \left({2\over 5(k+1)(k+3)} (h^k_2 - 30b^k)\right) \label{constraint1}\\
 {h'}^k &=& {16\over 15}h^k_2 \,.\label{constraint2}
\eea
As we will see below, the presence 
of these constraints 
will be especially 
important in the analysis of the extremal 3--point functions.

\subsection{Cubic action}
The kinetic term for the dilaton in the 10--dimensional action is
\be
S={1\over 2\kappa_{10}^2} \int d^{10}x\, \sqrt{G}\; {1\over 2} G^{mn}
\,\partial_m \phi \partial_n\phi \,.
\ee
The dilaton also occurs through its 
coupling to the two--form fields and to the axion, 
but  these latter  fields 
will not be excited in our analysis, and it is
consistent to ignore these terms in the action. We can also set
\be
h_{\m\a}\equiv 0\quad h_{(\a\b)}\equiv 0 \,.
\ee
Then we can expand 
\be
\sqrt{G}=\sqrt{g_1}\sqrt{g_2}
\left(1-{1\over 3}h_2+{1\over 2} h' +\dots \right)
\ee
where $g_1$, $g_2$ indicate the determinant of the background
metric on $AdS_5$ and $S^5$ respectively.
We obtain for the dilaton kinetic term the 
expansion (to cubic order in the fluctuations)
\bea \label{cubicaction}
S&=&{1\over 2\kappa_{10}^2}\int d^{10}x \, \sqrt{g}\; \left[{1\over 2}
D_\m\phi D^\m\phi+{1\over 2}
D_\a\phi D^\a\phi \right.\\&+&
\left.\left({1\over 4}h'-{4\over 15}h_2\right)D_\a\phi D^\a\phi+
{1\over 4}h'D_\m\phi D^\m\phi-{1\over 2} h'_{\m \n}D^\m\phi D^\n\phi\right]\,.
\nonumber
\eea

\subsection{Dimensional reduction}

We will encounter only scalar fields on the $S^5$, and we can 
expand these fields in scalar spherical harmonics. 
The spherical harmonics are normalised (following \cite{seiberg},
Appendix B) such that 
\bea\label{sphcont}
\int Y^{k_1} Y^{k_2}& = &z(k) \delta^{k_1k_2},\\
\int Y^{k_1} Y^{k_2} Y^{k_3} &          =& a(k_1,k_2,k_3) \cabc,
\label{pain}
\eea
where
\bea
z(k)&=& {1 \over 2^{k-1} (k+1)(k+2) }\\
a(k_1,k_2,k_3)&=&{\omega_5 \over ( \Sigma +2)! 2^{\Sigma -1}}
{k_1! \ k_2! \ k_3! \over \a_1! \ \a_2! \ \a_3!} \,.
\eea
Here $\a_1=\half (k_2+k_3-k_1)$, $\a_2=\half (k_1+k_3-k_2)$,
$\a_3=\half(k_1+k_2-k_3)$, $\Sigma=\half (k_1+k_2+k_3)$ and 
$\omega_5=\pi^3$ is the area of a unit 5-sphere.
Note that for notational simplicity we omit to indicate explicitly 
that each field is an element of 
a vector space of harmonics -- we  simply indicate the 
value of $k$.
We will also not explicitly write the group--theoretic 
factors  $\cabc$  in the equations below, 

To compute the 3--point functions  $ \< \O_t^{k_1}  \O_\phi^{k_2}  \O_\phi^{k_3}  \>$ we need to consider
 excitations where the fields $s^k$ are set to zero. 
Then from the definition (\ref{defs}) of $s^k$ and the constraints we find
\bea
 h_2^k& =& 10(k+4)b^k \\
h'_{\m\n}& =& h'_{(\m\n)} + {h' \over 5} g_{\m\n}= {2\over
5(k+3)(k+4)}D_\m D_\n h_2+ {2\over 15}{k\over k+3}h_2 g_{\mu\nu}\\
h_2^k& =&10(k+4)t^k
\eea
We then get for the dimensionally reduced form of the action 
(\ref{cubicaction})
\bea \label{dimredaction}
S&=&{1\over 2\kappa_{5}^2}\int d^{5}x \, \sqrt{g_{{1}}}\; \left[{z(k)\over 2}
\left(D_\m\phi^k D^\m\phi^k-k(k+4)
\phi^k \phi^k \right)\right.\\&+&
\left.a(k_1,k_2,k_3)\left(2{(k_1+4)^2\over (k_1+3)}\; 
t^{k_1}D_\m\phi^{k_2} D^\m\phi^{k_3}
-{2\over k_1+3}\;D_\m D_\n t^{k_1}D^\m\phi^{k_2}
D^\n\phi^{k_3}\right)\right]
\nonumber\eea
Note that the term $({1\over 4}h'-{4\over 15}h_2)$ is zero 
by the constraint equation (\ref{constraint2}), 
so that we have no `direct' 
coupling to the $h_2$ field; there is however an `indirect'
coupling given through the excitation of the field $h'_{\mu\nu}$.
The gravitational coupling constant in (\ref{dimredaction}) is related
to the SYM parameter $N$ by
\be \label{kappa}
2 \kappa_5^2 = \frac{8 \pi^2}{N^2}\,.
\ee

\subsection{Evaluation of the action}

The first cubic term in (\ref{dimredaction}) can be manipulated as follows
\bea
 &&\int_{\AdS_5} t^{k_1}D_\m\phi^{k_2} D^\m\phi^{k_3}=\int_{\AdS_5}
 t^{k_1} \,\half D^\m D_\m (\phi^{k_2}\phi^{k_3})-\half
(\mf{2}+\mf{3}) \,t^{k_1}\phi^{k_2}\phi^{k_3}\nonumber\\&&
=\int_{\AdS_5} \half (\mt{1}-\mf{2}-\mf{3})
\,t^{k_1}\phi^{k_2}\phi^{k_3}
+\half \int_{\partial(\AdS_5)} t^{k_1} D_n (\phi^{k_2} \phi^{k_3})
\nonumber\\&&-
\half \int_{\partial(\AdS_5)}\phi^{k_2} \phi^{k_3} D_nt^{k_1} \label{liu}
\eea
where we have used the equation of motion for the fields at the linear level.
$D_n$ indicates the outward normal derivative to the boundary and
we have introduced the symbols $\mf{}$ and $\mt{}$ to denote the masses of
the fields $\phi^k$, $t^k$ 
\be
\mf{}\equiv k(k+4)\qquad\mt{}\equiv (k+4)(k+8) \,.
\ee
We now observe that the boundary integrals 
found in the last step in (\ref{liu})
cannot contribute to the 3--point function 
if all the three points are disjoint\footnote{We thank
Oliver DeWolfe for this observation which also invalidates the form of 
extremal cubic interaction proposed in Appendix A of \cite{scalar}.}. 
This is the case because 
\be
\int_{\partial(\AdS_5)} t^{k_1} D_n\phi^{k_2} \phi^{k_3} = 
\int \int_{\partial(\AdS_5)} t^{k_1}(x) {\partial \over \partial z_n}
K(z,x')\phi^{k_2}(x') \phi^{k_3}(x)dx dx'
\ee
where $K$ is the propagator from a boundary point 
$x'$ to a bulk point $z$, 
and it is assumed in the above equation that we take the 
limit $z\rightarrow x$. But now note that if we vary such a contribution to the 
action with respect to the boundary values of the fields, we will obtain a
nonzero value only if two of the points where we consider field variations are
coincident (namely, t he points where the fields $t^{k_1}$ and $\phi^{k_3}$ are
varied).  We will be interested only in the values of the correlators  for
separated points. Thus terms of the form of the boundary integrals in (\ref{liu})
will always be dropped.

To analyse the last term in (\ref{dimredaction}) we carry out the following steps. We first define
\be
P_{\m\n}=\half\left(D_\m \phi^{k_2} D_\n \phi^{k_3}+D_\n \phi^{k_2}
 D_\m \phi^{k_3}\right)-\half g_{\m\n} D^\lambda \phi^{k_2} D_\lambda \phi^{k_3}
\ee
which satisfies the relation
\be
D^\m P_{\m\n}=\half\left(\mf{3}\phi^{k_3} D_\n \phi^{k_2}+\mf{2}\phi^{k_2} D_\n
 \phi^{k_3}\right)\,.
\ee
Then we find
\bea
\int_{\AdS_5}\;D_\m D_\n\, t^{k_1}D^\m\phi^{k_2}
D^\n\phi^{k_3}=\int_{\AdS_5}D_\m D_\n\, t^{k_1}\left(P_{\m\n}+\half g_{\m\n}
D^\lambda \phi^{k_2} D_\lambda \phi^{k_3}\right)
\nonumber~~~~~~~~~~~~~~~~~\\
=\half\int_{\AdS_5}\, \left[\mt{1}t^{k_1}D^\lambda \phi^{k_2}
D_\lambda \phi^{k_3}-D^\m\,t^{k_1}\left(\mf{2}\phi^{k_2}D_\m \phi^{k_3}
+\mf{3}\phi^{k_3}D_\m
\phi^{k_2}\right)\right]
\nonumber\\
+\int_{\partial(\AdS_5)}\,D^\m\,t^{k_1}P_{\m n}
~~~~~~~~~~~~~~~~~~~~~~~~~~~~~~~~~~~~~~~~~~~~~~~~~
\,.
\eea
The above expression can be further manipulated by steps similar to those in 
(\ref{liu}). Collecting all contributions, we get
\bea
&&2\kappa_{5}^2\,S_{\rm cubic}
={{\int_{\AdS_5}\,a(k_1,k_2,k_3)\,t^{k_1}\phi^{k_2}\phi^{k_3}}}\\
&&\cdot {\textstyle{
\left[{(k_1+4)^2\over
k_1+3}\left(\mt{1}-\mf{2}-\mf{3}\right)+{1\over
2(k_1+3)}\left((\mf{2}-\mf{3})^2-m^4_t(k_1)\right)\right]}}+
\nonumber\\
&&\int_{\partial(\AdS_5)}\,{\textstyle{{a(k_1,k_2,k_3)\over k_1+3}}}\,
\left(-D_n\phi^{k_3} D_\m\,t^{k_1}D^\m \phi^{k_2}-D_n\phi^{k_2} D_\m\,t^{k_1}D^\m
\phi^{k_3}+D_n\,t^{k_1}D^\lambda \phi^{k_2} D_\lambda \phi^{k_3}
\right)\nonumber\eea Using the explicit expressions for the masses of the fields
we finally get
\bea \label{surface}
&& 2\kappa_{5}^2\,S_{\rm cubic}\label{finalaction}
= -8\frac {\left (\Sigma+4\right )
\alpha_1\left (\alpha_2+2\right )
\left (\alpha_3+2\right
)}{(k_1+3)}{{\int_{\AdS_5}\,a(k_1,k_2,k_3)\,t^{k_1}\phi^{k_2}\phi^{k_3}}}\;+
 \\
&& \int_{\partial(\AdS_5)}\,{\textstyle{{a(k_1,k_2,k_3)\over k_1+3}}}\,
\left(-D_n\phi^{k_3} D_\m\,t^{k_1}D^\m \phi^{k_2}-D_n\phi^{k_2} D_\m\,t^{k_1}D^\m
\phi^{k_3}+D_n\,t^{k_1}D^\lambda \phi^{k_2} D_\lambda \phi^{k_3} \right)\nonumber
\eea

\subsection{Computing the 3--point function}

There are two kinds of terms in (\ref{finalaction}) -- a bulk vertex and 
a boundary term. 
We imagine carrying out the evaluation of the action in some large 
but finite region of AdS space, 
with the values of the fields fixed at the boundary of this region. 
Varying the value of the action with respect to the appropriate 
fields at chosen points on the boundary 
we will obtain the 3--point function.

We will find that the case  $k_1<k_2+k_3$ and the case $k_1=k_2+k_3$ 
(the `extremal' case) need to be analyzed somewhat differently. 
In the first case, 
it is easy to see that the boundary integral vanishes as we take 
the size of the AdS region to infinity, faster than the expected scaling of 
the three point function. The bulk integral gives a nonvanishing result, 
which scales exactly as expected for the 3--point 
function of the boundary CFT. 

In the extremal case, on the other hand, we find that the coefficient of the 
bulk integral is zero. 
The boundary term in this case gives a nonzero 
contribution however, since it scales in exactly the manner expected of 
the 3--point function. Evaluating this boundary integral needs some care, 
and we will carry out the evaluation below.

\subsubsection{The three--point function for $k_1<k_2+k_3$}

We define  $\O_\phi^k$ and $\O_t^k$ 
as the operators that couple to the boundary value of 
the supergravity fields
 $\phi^k$ and  $t^k$ with strength 1, {\it i.e.} we let
the source terms in the action  be
\be
\int d^4 x \; \phi^k(x) \O_\phi^k (x) ,\qquad  \quad
\int d^4 x \; t^k(x) \O_t^k (x) \,.
\ee
The bulk integral over $AdS_5$ in (\ref{finalaction}) 
was evaluated in \cite{april}, equation (25),
for fields of generic dimensions $\Delta_i$. 
 It suffices
to note that the conformal dimensions $\Delta_i$
of $t^{k_1}$, $\phi^{k_2}$ and $\phi^{k_3}$ are  related to the KK levels
$k_i$ through
\be
\Delta_1 =k_1+8 \quad \Delta_2=k_2+4 \quad \Delta_3=k_3+4 \,.
\ee
Then we find that the three--point function is
\bea
&&\< \O_t^{k_1} (x_1) \O_\phi^{k_2} (x_2) \O_\phi^{k_3} (x_3) \>=
\frac{1}{2 \kappa_5^2} \frac{4}{\pi ^4} \, 
\frac{a(k_1,k_2,k_3)}{x_{12}^{8+2\alpha_3}
x_{13}^{8+2\alpha_2}x_{23}^{2\alpha_1}}
\\ && \qquad \times 
\frac {\left (\Sigma+4 \right )\left (\alpha_2+2\right )\left (
\alpha_3+2 \right )}{\left ({k_1}+3\right )}
\,\frac {
\Gamma (\alpha_1+1)\Gamma (\alpha_3+4)\Gamma (\alpha_2+4)\Gamma
(\Sigma+6)
}
{ 
\Gamma ({k_1}+6)
\Gamma ({k_2}+2)\Gamma ({k_3}+2)
} \nonumber
\eea
Although the above calculation is 
strictly valid only for the range $k_1<k_2+k_3$,
we observe that there is a smooth limit
as $k_1 \rightarrow k_2 +k_3$:
\bea \label{limit}
&&\lim_{k_1 \rightarrow k_2+k_3}
\< \O_t^{k_1} (x_1) \O_\phi^{k_2} (x_2) \O_\phi^{k_3} (x_3) \> 
=\frac{1}{2 \kappa_5^2}\, {4\,\over \pi^4}
\frac{a(k_2+k_3,k_2,k_3)}{x_{12}^{8+2\,k_2} x_{13}^{8+2\,k_3}}
\\ && ~~~~~~~~~~~~~~~~~~~
\times
{\frac {\left ({k_2}+3\right )\left ({k_2}+2\right )^{2}
\left ({k_3}+2\right )\left ({k_3}+3\right )^{2}
\left ({k_2}+{k_3}+4\right )}{\left ({k_2}+{k_3}+3\right )}}\nonumber \,.
\eea

\subsubsection{The extremal case $k_1=k_2+k_3$}

In this case the coefficient of the bulk integral in (\ref{finalaction})
vanishes. The contribution comes from the boundary term, which must
be regulated and calculated carefully. We introduce a cutoff at 
$z_0 = \epsilon$, where $z_0,z_i$ are the coordinates of the usual 
upper half-plane metric for AdS, and require that bulk fields satisfy
a Dirichlet boundary value problem there. This is a geometrically
well defined prescription, and leads to the free solution
\be
t^k(z) = \int d^4x\ K^\epsilon_\Delta (z,x) \bar t^k(x)
\ee
where $K^\epsilon_\Delta (z,x)$ is the Poisson/Dirichlet kernel for the
cutoff space-time and $\bar t^k(x)$ is the boundary source for $t^k(z)$.
A similar equation holds for $\phi^k(z)$. We must insert the free solution
into \ref{dimredaction} or \ref{finalaction} and obtain the contribution
to the action functional which is cubic order in the sources.

A certain difficulty now emerges. Both bulk integrals in
 (\ref{dimredaction}) are convergent for regular boundary data, i.e. smeared
sources, but the integral from the interaction vertex with four
derivatives diverges for point sources, 
i.e. $\bar t^k(x) \sim \delta^4 (x-\hat x)$. The divergence comes from the
region where $z$ is very close to the insertion point $\hat x$ where
$K^\epsilon_\Delta (z,x)$ can be well approximated by its flat space form
\be (\label{kflat})
{(z_0-\epsilon)\over [(z_0-\epsilon)^2+(z-x)^2]^{5/2}} \, .
\ee
This divergence for point sources is an artefact of the cutoff procedure
which would occur for all values of the $k_i$, not just extremal cases.

In the Appendix the behavior of the singular integral is studied for 
smeared sources
which are peaked about the final boundary insertion points. It is shown
that the point source limit is well defined, i.e. finite and
independent of details of the smearing. In particular this means that the
contribution to the integral of a small strip $\epsilon <z_0 < \delta$
vanishes as $\delta \rightarrow 0$. 

While the Appendix explains that the integral is well defined, a position 
space approach does not give an easy way to calculate
the 3-point function. Instead we used the Fourier transform, since
plane waves may be viewed as particular choices of smeared sources for
which both the bulk integrals of \ref{dimredaction} converge. For example
the Fourier transform of $K^\epsilon_\Delta (z,x)$ is well known to be
\cite{polyakov,april}
\be
K^\epsilon_\Delta(p)= \frac{
z_0^{d \over 2} \,{\cal K}_{\Delta-\frac{d}{2}}
(p z_0) }
{
\epsilon^{d \over2} \,{\cal K}_{\Delta-\frac{d}{2}}
(p \epsilon)}
\ee
where ${\cal K}_\nu$ is the modified Bessel function of index $\nu$, and
$d$ is the boundary dimension of $AdS_{d+1}$. (We will specialize to $d=4$
in our final result for the correlator.) We see that $K^\epsilon_\Delta(p)$
admits derivatives of arbitrary order with respect to $z_0$ which have
smooth limits to the boundary at $z_0=\epsilon$.

The partial integrations of Section 3.5 are valid with plane wave sources,
so that in the end it is just the Fourier transform of the boundary
interaction in (\ref{finalaction}) that must be calculated.  
It is easy to see that the derivatives in (\ref{finalaction}) which
are parallel to the boundary vanish faster in the eventual limit
$\epsilon \rightarrow 0$ than those in the normal direction, so we
keep only the dominant $z_0$ derivatives. The 3-point correlator in
momentum space is 
just a product of three $K^\epsilon_\Delta(p)$ with normal derivatives 
applied. 

We use the asymptotic formula
\be
D_n K^\epsilon_\Delta(p) =
z_0 \frac{\partial}{\partial z_0} K^\epsilon_\Delta(p) \bigg |_{ z_0
=\epsilon}=
(d-\Delta) +\dots a_\Delta (p \epsilon)^{2(\Delta-\frac{d}{2})} ln(p\epsilon)
  +\dots 
\ee
where the first $\dots$ refer to positive integer powers of $p$ and 
the second $\dots$ to terms containing $ln(p\epsilon)$ times higher powers
of $p$.
The coefficient $a_\Delta$ is easily obtained from standard treatments
of Bessel functions, but for our purposes it suffices to recall
that $ a_\Delta \,p ^{2(\Delta-\frac{d}{2})}$ is the 
Fourier transform of the correctly normalized
expression of the 2--point function 
(Appendix of \cite{april}), which is
\be \label{2pt}
\frac{1}{(x-y)^{2 \Delta}} \frac{(2 \Delta-d)\Gamma(\Delta)}
{\pi^{\frac{d}{2}} \Gamma(\Delta-\frac{d}{2})} \,.
\ee

The relevant term in the product of three propagators is
\bea \label{3K}
D_n K^\epsilon_{\Delta_2+\Delta_3}(p_1)D_n K^\epsilon_{\Delta_2}(p_2)
D_n K^\epsilon_{\Delta_3}(p_3) ~~~~~~~~~~~~~~~~~~~~~~~~~~~~~~~\nonumber \\
=
\dots + \epsilon^{2 \Delta_2 + 2 \Delta_3-2d}
(d-\Delta_2-\Delta_3) a_{\Delta_2} a_{\Delta_3} p_2^{\Delta_2-\frac{d}{2}}
ln(p_2)p_3^{\Delta_3 - \frac{d}{2}}ln(p_3)+\dots
\eea
As in previous $p$-space calculations of 2-point functions we have kept 
the leading term as $\epsilon \rightarrow 0$
which is non--analytic in both $p_2$ and $p_3$. 
An interesting feature of this term is that it
depends on $p_2$ and $p_3$ in a factorized way:
it is in fact the Fourier transform of a product of two 2--point
functions.
The factor $\epsilon^{2 \Delta_2 + 2 \Delta_3-2d}$
provides the correct scaling behavior 
(namely, $\O(\epsilon^{\Delta_1+\Delta_2+\Delta_3-2d})$)
for a 3--point function
of operators of dimension  $\Delta_2$,
$\Delta_3$, $\Delta_1=\Delta_2+\Delta_3$.
Observe
that in the non--extremal cases $\Delta_1<\Delta_2 +\Delta_3$
the boundary term (\ref{3K}) scales too fast to give a contribution. 

${}$ From (\ref{3K}) and (\ref{2pt}), taking into account
the overall coefficient in the boundary term of the action
(\ref{finalaction}),
we finally get for the 3--point function
\bea \label{3ptextremal}
&& \< \O_t^{k_2+k_3} (x_1) \O_\phi^{k_2} (x_2) \O_\phi^{k_3} (x_3) \>
=  \frac{1}{2 \kappa_5^2} {4 \over \pi ^4}
\frac{a(k_2+k_3,k_2,k_3)}{x_{12}^{8+2\,k_2} x_{13}^{8+2\,k_3}}
\nonumber \\
&& ~~~~~~~~~~~~~~\times
{\frac {\left ({k_2}+3\right ) \left ({k_2}+2\right )^{2}
\left ({k_3}+3\right ) \left ({k_3}+2\right )^{2}
\left ({k_2}+{k_3}+4\right )}{\left ({k_2}+{k_3}+3\right )}}\nonumber\,.
\eea 
Comparison with the
expression (\ref{limit}) shows exact agreement.

One might have thought that a simpler way to calculate the extremal
3-point function is to calculate the two bulk integrals in
(\ref{dimredaction}) using a simple cutoff at $z_0=\epsilon$
with standard bulk--to--boundary propagators \cite{witten}
\be
K_\Delta (z,x) \sim
\left({z_0\over {z_0^2 +(z -x)^2} }\right)^\Delta\,.
\ee
Each integral is logarithmically divergent, but the divergence cancels
in the sum. The final result depends on how the integrals are cut off,
and we discuss two methods. First one can
continue $k_1$ into the convergent region $k_1<k_2+k_3$ so that each
integral contains a pole $1/(k_2+k_3-k_1)$ which cancels between them.
The final result agrees with (3.48) for non-extremal correlators and its
continuation to the extremal case agrees with (3.42). This tests the
equivalence of the actions (\ref{dimredaction}) and (\ref{finalaction}) in
the non-extremal region, but it does not give an independent evaluation
of the extremal case. The second method is to cut off each divergent
integral at $z_0=\epsilon$, do the integrals over the 4 coordinates $z_i$,
and then observe that the divergent part near $z_0=0$ cancels in the 
integrand when both contributions are combined. The final result then
does not agree with the method of analytic continuation. However,
the situation of two cancelling divergent integrals is very similar to
that of 2-point functions \cite{witten,april,klebwit}, and we
believe that it is incorrect to use a simple cutoff without imposing
Dirichlet boundary conditions. In the case of 2-point functions it was
shown \cite{april} that the supergravity calculation of a 3-point function
of a current and two scalar operators is unambiguous. The Ward identity
then gave a scalar 2-point function which agreed with the calculation
by the Dirichlet method. In the present case we suggest that the 4-point
function of a current and an extremal combination of scalar operators will
also be unambiguous, and the Ward identity will lead to an extremal
3-point correlator which agrees with (\ref{3ptextremal}).

\subsection{Discussion}

The agreement of extremal 3--point functions calculated by the two methods
of analytic continuation in the scale dimensions $k_i$ and carefully 
regulated boundary interaction is the principal result of this section. 
The result was obtained in a descendent sector of the theory because it
was technically easier to find the interaction Lagrangian. However,
the 3-point correlators of chiral primary
operators and their descendents are related by $\N=4$ supersymmetry
transformations, so our computations also justify the previous
calculation of $\< \O_{k_1}\O_{k_2}\O_{k_3}\>$ obtained in \cite{seiberg}
by (implicit) analytic continuation. The agreement with free field
theory results then shows that the fields $s^k$ used in \cite{seiberg}
couple only to the single--trace operators $\O_k$ rather than to
admixtures with protected multi--trace operators. A 3-point correlator
of one double-trace and two single-trace operators
can then be obtained within the $AdS_5/CFT_4$ correspondence by taking a
suitable short distance limit of a 4-point correlators of single traces.

Our calculations involved supergravity fields analogous to the $s^k$ of
\cite{seiberg}, and we chose to process the derivative interactions in
(\ref{dimredaction}) by partial integration rather than eliminate them
by field redefinition of the schematic form 
$s= s' + s'^2 + (\partial_{\mu} s')^2$ as was done in \cite{seiberg}. We
have considered the question of a similar transformation of the fields
$t^k$ and $\phi^k$. For example, for the lowest harmonics
$k_1=k_2=k_3=0$, which is an extremal configuration, the transformation
\bea
t'^0=t^0+{a(0,0,0)\over 3}D_\mu\phi^0\,D^\mu\phi^0 \nonumber\\
\phi'^0=\phi^0+{2a(0,0,0)\over 3}D_\mu t^0\,D^\mu\phi^0.
\eea
takes us directly from (\ref{dimredaction}) to an  action  for $t'^0$ and
$\phi'^0$ which is free through cubic couplings. The equations of motion are not
used in this transformation and no boundary terms appear.
  One might be tempted to say that the new fields couple to
mixtures of single-- and double--trace operators (as proposed in
\cite{russians} for $s'^k$ fields), 
but we could not 
demonstrate this to our satisfaction. It may also be the case that
the required transformation of fields is inadmissable because it is 
non--invertible or not
compatible with the boundary value problem used in the $AdS/CFT$ 
correspondence. Frankly, we are still confused about the role
of the $s'^k$ fields, and we suggest that it is an interesting issue
for future study.

\setcounter{equation}{0}
\setcounter{footnote}{0}

\section{Non--renormalization conjectures for extremal n--point functions}

Although the considerations of this Section apply to all extremal
$n$--point functions of chiral primary operators $\O_k$, we discuss
the simplest case of 4--point functions in detail and then briefly
indicate the general line of argument. We consider correlators 
$\<\O_{k_1}  \O_{k_2} \O_{k_3}\O_{k_4}\>$, with $k_1 = k_2 +k_3 +k_4$,
as they are computed in the large $N$ limit 
at strong `t Hooft coupling
from the $AdS_5/CFT_4$
correspondence. We shall now argue that:

i) The structure of the Type IIB supergravity action requires that 
   the space--time form of these correlators is a product of 2--point
   functions, specifically
\be
\<\O_{k_1}(x)  \O_{k_2}(y) \O_{k_3}(z)\O_{k_4}(w)\> = 
 \frac{A}{(x-y)^{2k_2}(x-z)^{2k_3}(x-w)^{2k_4}}
\ee
   where the numerical constant A depends on the scale parameters $k_i$.
This is the same functional form as in the free--field approximation.

ii) Compatibility of this factored form with operator product expansions
    requires that the constant $A$ is not renormalized.

Thus  for large $N$, the extremal 4--point functions
take the same value at large `t Hooft coupling $\lambda= g_{YM}^2 N$
as  in the free--field approximation.
It is natural to conjecture that {\it the extremal 4--point functions of
chiral primary operators at large $N$ are independent of
$\lambda=g_{YM}^2N$.} As for 3--point functions
\cite{seiberg,dhfskiba} a stronger version of the conjecture
is that these 4--point functions are independent of $g_{YM}$ for any
$N$.

We will take the point of view that the result for the extremal 
correlator is correctly given by the procedure of analytic
continuation in the conformal dimensions $k_i$. The result
of the previous Section makes us confident that this is indeed
the case. It should be possible to recover
all of the following analysis by working at exact extremality
and carefully considering the boundary--like interactions studied
in the previous Section,
and we will make some remarks on how we expect this to happen.

We need to consider two types of contributions to
$\<\O_{k_1}\O_{k_2} \O_{k_3} \O_{k_4}\>$: exchange diagrams with two 3--point
couplings, and a quartic graph with one 4--point coupling.

Let us begin by examining the exchange diagrams.
Without loss of generality we can
consider the case where $s_{k_1}(x_1)$ and $s_{k_2}(x_2)$ join by a cubic
vertex to some  intermediate field $\phi$, 
which then joins by a cubic vertex to
the other two fields at $x_3$ and $x_4$. Each of the fields $s_{k_i}$
is in $SU(4)$ representation with Dynkin label $(0,k_i,0)$. Multiplying
the representations of the fields at $x_1$ and $x_2$ and
also taking the product of representations for the other pair, is easy to see
that the only common representation between these two products has Dynkin
label $(0,k_3+k_4,0)$. 
This must be the representation of the
intermediate field $\phi$, and there are two possibilities:
either $\phi$ is the primary field $s_{k_3+k_4}$ or it is
a $SU(2,2|4)$  descendent. By detailed consultation of the
Tables in \cite{vannieuv} which present the spectrum of the dimensionally
reduced Type IIB theory and those of \cite{gunmarc} on the structure of
the relevant representation of the superalgebra $SU(2,2|4)$, one learns
that the possible descendent states in the same $SU(4)$ representation
have dimension $\Delta> k_3+k_4$ and they are superconformal descendents
of chiral primaries of dimension $\tilde k > k_3+k_4$. We now proceed to
discuss in turn the exchange diagrams for primaries and descendents.

(a)\quad If the exchanged field $\phi$ is the chiral primary $s_k$ of dimension
$k=k_3+k_4$, then we must consider the cubic couplings
${\cal G}(k_1,k_2,k)s_{k_1} s_{k_2} s_k$ and 
${\cal G}(k_3,k_4,k)s_{k_3} s_{k_4} s_k$. However both vertices are
extremal and the coupling constants vanish. Stated in terms of a calculation
of the exchange diagram by analytic continuation, we find a double zero in 
the numerator. However, each of the two integrals over $AdS_5$ produces a
pole so the net amplitude is finite. One way to see this is to generalize
the argument used for the contact graph. In the first step one sees that the
integral diverges when the adjacent internal vertex approaches the boundary
point $\vec x_1$ of the highest dimension operator. The singular pole factor
$1/\delta$ of (4.2) multiplies the space--time product of 
$1/(\vec x_1-\vec x_2)^{2k_2}$ times a further divergent integral which is 
exactly the extremal 3-point function of primary operators of dimension
$k_3,k_4$ and $k_3+k_4$. The net result for the amplitude is a finite
multiple of the factorized form of (\ref{product}). One can also see that the same
result is obtained from the scalar exchange integral computed in Sec. 3c of 
\cite{scalar} which has a double pole for the relevant values of scale
dimensions of external and exchanged fields. 

(b) \quad The treatment of descendent exchange graphs is more complicated.
First note  \cite{vannieuv} that the descendents in question are
scalars and symmetric  tensor fields in the
supergravity theory, so higher spin exchange diagrams which have not yet
been studied in general form are involved. The coupling constants for the
cubic vertices $\phi s_{k_1} s_{k_2}$ and $\phi s_{k_3} s_{k_4}$ are
related by supersymmetry to the primary vertices\footnote{Recently,
all the cubic couplings between two $s$ fields and any
other supergravity field have been explicitly determined \cite{russians,lee}.}
 in the previous paragraph
except that now we have $\tilde k>k_3+k_4$. The coupling 
${\cal G}(k_3,k_4,\tilde k)$ vanishes (by $SU(4)$ flavor symmetry), but 
${\cal G}(k_1,k_2,\tilde k)$ does not vanish. 
Thus only one of the two descendent couplings vanishes. Consider the case of an
exchanged scalar descendent field. It can be seen from examination of Sec 3c of
\cite{scalar} that the exchange integral has a single pole for relevant values of
the dimensions, and that the singular factor multiplies exactly the space--time
function we need. The divergence in this case comes when both interaction
vertices are within a small region near  the boundary point $\vec x_1$ of the
highest dimension operator.  The singular contribution does not depend on the
spin of the exchanged  field, since the short distance behavior of the
bulk--to--bulk propagator is universal.  Thus we believe that both descendent
scalar and tensor exchange  graphs contribute finite multiples ofthe desired
factored space-time form (\ref{product}).

Next, we analyse the quartic graph, which (after an appropriate field
redefinition) is obtained from a 4--point vertex of the form 
$$ \int_{AdS_5} \,{\cal G}(k_1,k_2,k_3,k_4) s_{k_1} 
s_{k_2} s_{k_3} s_{k_4}\, .
$$
The coefficient ${\cal G}(k_1,k_2,k_3,k_4)$ of this vertex in the supergravity
action is not yet known,  although it is certainly possible to obtain it by
extending the analysis of \cite{seiberg}. 

We now observe that the integral over the $AdS_5$ space involved 
in the extremal quartic graph diverges. 
The divergence comes from the region where the
integration variable approaches the location of the operator 
of higher dimension $\O_{k_1}$ on the boundary. 
To see this divergence, note that near this point the
propagator $K_{k_1}$ behaves as $({z_0\over (z-\vec x_1)^2})^{k_1}$, while  each 
of the other propagators behaves as $(z_0)^{k_i}$, $i=2,3,4$.
The integration measure is ${d^5z\over z_0^5}$. Thus we see that the integral is
logarithmically divergent at $z\rightarrow \vec x_1$. 

Assembling the results for the quartic term with those of the exchange 
diagrams, for the case of extremal correlators, we conclude the following.
Since the exchange diagrams yield finite contributions, and the AdS integral 
for the quartic diagram diverges, a finite result for the 4--point function
requires  that the quartic coupling vanishes at extremality.
All of the following considerations are subject to this plausible
(and in principle checkable) assumption.
 
To regulate the divergent integral, and find the value of the extremal 4--point
function, consider the following procedure of analytic continuation.
Let the fields $s_{k_i}$, for $i=2,3,4$,  have dimension $k_i+\delta$, while we
keep the dimension of $s_{k_1}$ to be $k_1$. 
The integral is then finite, though it
is still dominated by the region in the infinitesimal vicinity of
$z\rightarrow \vec x_1$. Thus the propagators $K_{k_i}$, 
for $i=2,3,4$,  be approximated by the form 
$$
{z_0^{k_i+\delta}\over (\vec x_i-\vec x_1)^{2\,k_i}}\, .
$$
The integral thus gives
\be
\prod_{i=2,3,4}{1\over (\vec x_i-\vec x_1)^{2\,k_i}}\;\int {d^5z\over z_0^5}\;
z_0^{k_1+3\delta}\left({z_0\over (z-\vec x_1)^2}\right)^{k_1}\sim 
\prod_{i=2,3,4}{1\over
(\vec x_i-\vec x_1)^{2\,k_i}}\;{1\over \delta}
\ee
In line with our assumption that the coupling is zero
at exact extremalty,  and in analogy with the 3--point case,
it is plausible that if we analytically continue the
dimensions as above  the coupling
${\cal G}(k_1, k_2 +\delta, k_3 + \delta, k_4+\delta)
\sim \delta$. Then the
quartic graph gives
\bea\label{product}
\sim \delta\;{1\over \delta}\prod_{i=2,3,4}{1\over
(\vec x_i-\vec x_1)^{2\,k_i}} \sim
\<\O_{k_2}(\vec x_2)\O_{k_2}(\vec x_1)\>\<\O_{k_3}(\vec x_3)\O_{k_3}(\vec x_1)\>
\<\O_{k_4}(\vec x_4)\O_{k_4}(\vec x_1)\>
\eea
which is exactly the factorized form of (4.1).

Instead of the analytic continuation procedure just outlined,
it should be possible to carry out a careful analysis at exact extremality.
The bulk coupling ${\cal G}(k_1,k_2,k_3,k_4)$ vanishes
at extremality, but we expect that in the
dimensional reduction a boundary interaction of
the form $\int_{{\rm \partial(AdS_5)}} D_n s_{k_1} D_n s_{k_2}
D_n s_{k_3}D_n s_{k_4}$ is induced through the constraints. 
It is easy to check
(for example by the momentum space analysis of the previous Section)
that such an interaction reproduces the factorized
functional form (\ref{product}). (Note that as in the case
of 3--point functions, this boundary term gives a contribution only 
at exact extremality, $k_1=k_2+k_3+k_4$).

We conclude that the large $N$, strong `t Hooft coupling limit
of the extremal 4--point functions of chiral primary operators,
computed in supergravity,
takes the factorized form of a product of three 2--point functions,
as in (\ref{product}).
This is the same functional form as in the free field theory
approximation, and it thus suggests that extremal 4--point functions
are not renormalized; that is they are independent of $\lambda =g_{YM}^2N$,
at least for large 
$N$! Assuming the factorized form it is quite remarkable that one
can prove non-renormalization
by an operator product argument similar to the argument at the beginning
of Section 2. The argument applies to the general extremal correlator
$\<\O_{k_1}  \O_{k_2} \O_{k_3} \O_{k_4}\>$, but to simplify the notation
we consider the particular case $k_1=6, k_2=k_3=k_4=2$ which we assume
has the form
\be\label{A}
\<\O_6(x)  \O_2(y) \O_2(z) \O_2(w)\> =
  { A\over (x-y)^4(x-z)^4(x-w)^4}.
\ee
We will show that $A$ is not renormalized, if the 2-- and 3--point functions
that appear in the argument below are not renormalized.

Consider the operator product expansion $\O_6(x)  \O_2(y)$ which contains
\be\label{62ope}
\O_6(x)  \O_2(y) \stackrel{y\rightarrow x}{\longrightarrow} c'_{624} {\O_4(x)\over (x-y)^4} + c'_{62(22)} {[\O_2(x)\O_2(x)]_
{\rm max}\over (x-y)^4}  +\dots
\ee
where $\dots$ indicates the contribution of operators in representations of
$SU(4)$ different from the common representation $(0,4,0)$ in the $\O_6\O_2$
and $\O_2 \O_2$ direct products, and also
operators in the $(0,4,0)$ representation belonging to long
multiplets.

The assumed factorized form of the 4--point function (\ref{A}) requires that 
the only common operators in the $\O_6\O_2$
and $\O_2\O_2$ $OPE$'s are of dimension 4 (and their derivative descendents
which are omitted for simplicity). This rules out a possible 
contribution to the double OPE from non--protected long--multiplet operators.  

We also need the forms of several 2- and 3-point functions, namely
\bea
\<\O_4(x) \O_4(y)\> &=& \frac{a_{44}}{(x-y)^8},\\
\<\O_4(x) [\O_2(y)\O_2(y)]_{\rm max}\> &=& {a_{4(22)}\over (x-y)^8},\\
\<[\O_2(x)\O_2(x)]_{\rm max}  [\O_2(y)\O_2(y)]_{\rm max}\> &=&
    {a_{(22)(22)}\over (x-y)^8}.\\
\<\O_4(x)  \O_2(y) \O_2(z)\> &=&{c_{422}\over (x-y)^4(x-z)^4},\label{422}\\
\<\O_6(x)  \O_2(y) \O_4(z)\> &=& {c_{624}\over(x-y)^4(x-z)^8},\label{624}\\
\<\O_6(x)  \O_2(y) [\O_2(z) \O_2(z)]_{\rm max}\> &=&
     {c_{62(22)}\over (x-y)^4(x-z)^8}.\label{62(22)}
\eea
All coefficients above are assumed to be non--renormalized as a consequence
of the arguments of \cite{seiberg,dhfskiba,skibanew}, and the argument of
Sec 2 shows that $a_{4(22)}= c_{422}$. Using the OPE (\ref{62ope}) in
the 3-point functions (\ref{624}) and (\ref{62(22)})  one finds the relations
\bea
c'_{624} a_{44} + c'_{62(22)} a_{4(22)} = c_{624},\\
c'_{624} a_{4(22)} + c'_{62(22)} a_{(22)(22)} = c_{62(22)}.
\eea
which imply that the OPE coefficients $c'_{624}$ and $c'_{62(22)}$ are
not renormalized.

The final step is to take the limit $y\rightarrow x$ of the 4--point function (\ref{A})
and use (\ref{422}) to obtain
\be
A = c'_{624} c_{422} + c'_{62(22)} c_{22(22)},
\ee
which shows that $A$ is not renormalized.

These considerations naturally extend to higher point functions.
For extremal $n$--point functions of chiral primaries, all
disconnected diagrams necessarily vanish, since they factorize
into several connected graphs each of which cannot be an $SU(4)$ singlet.
Using arguments similar to the 4--point function case, one finds that 
connected diagrams yield, after the analytic continuation procedure, a 
factorized product of $(n-1)$ two--point functions. 
Therefore, we are led to the same non--renormalization
conjectures for the extremal $n$--point functions of chiral primaries
for any $n$.

\section{Appendix}

In this Appendix we present arguments that the extremal 3-point functions,
evaluated directly from the bulk version of the Type IIB supergravity action 
of (\ref{dimredaction}), are uniquely defined and finite. This method 
of evaluating the extremal 3-point functions is equivalent to that
starting from the pure surface action of (\ref{surface}), but it circumvents 
subtle divergences encountered at intermediate stages when dealing
directly with the surface action (\ref{surface}). 

To exhibit these subtle divergences, we start from the surface action
(\ref{surface}), which is all that remains at extremality.
In evaluating the 3-point functions, we encounter contributions of
the type
\be
\int_{\partial AdS} \partial_{z_0} K(z_0,x;x_1)\partial_{z_0}
K(z_0,x;x_2)\partial_{z_0} K(z_0,x;x_3) d^4x\, .
\ee 
The 4-dimensional $x$ integral above has singularities at $x=x_i$, $i=1,2,3$.   
It is not immediately clear how these singularities should be regulated. 
If the total 3-point function is to be finite, we should probably expect 
to find Dirac $\delta$-function contributions localized at the points
$x_i$ which can compensate for the divergences at $x=x_i$. Clearly, the 
regularization of both the surface integral and the $\delta$-functions
will be subtle and must be handled with great care. In particular, the
presence of the $\delta$-functions will require us to work with smeared out
sources.

What we show below is that the bulk integral (\ref{dimredaction}) that we start
with is itself  well defined. We do this by showing  that there is no divergence
in such an integral from the vicinity of the points $x_i$, the rest of the
integral is easily seen to be convergent. Having proved this convergence we see
that we can truncate the domain of integration of the bulk point $z$ slightly
above the boundary. The boundary is at  $z_0=\epsilon$, and so the bulk integral
will be restricted to $z_0>\epsilon+\delta$, with the limit
$\delta\rightarrow 0$ being taken before the limit $\epsilon\rightarrow 0$.
Since the bulk integral is convergent, the contribution to the integral
from the region $\epsilon<z-0<\epsilon+\delta$ is ignorable, but this
regularisation allows a sensible and well defined integration by parts. The
integration variable $x$ on the boundary is now at $z_0=\epsilon+\delta$,
and so does not collide with the source insertions at $x_i$ which are at a
value $z_0=\epsilon$. As mentioned in section 3, the result computed for
extremal correlators in this manner agrees with the analytic continuation
of the 3-point function in dimensions from the case where the correlator is
not extremal.

To see the regularity of the bulk integrals, consider the integrals cubic 
in the fields in (\ref{dimredaction}).  It is not hard to check that the first of
these integrals is convergent, so we will make some remaks about the second
integral, which help towards showing that this latter integral is
convergent too. The most significant divergence comes from the integral in
the vicinity of the point $x_1$, so we work in a ball around $x_1$, with
$x_2$ and $x_3$ assumed to be far away.

(1) \quad  We start with a regularisation where we have taken the boundary
of our space to be at $z_0=\epsilon$ rather than at $z_0=0$. As noted in
section 3, at short distances (less than the AdS curvature length scale) we
must use the boundary to bulk propagator appropriate to flat space, as given in
(\ref{kflat}).
Now we note that to derive the 3-point function, we must take a functional 
derivative with respect to the boundary values of the fields. To avoid
singularities associated with a delta function source, we start with a
source that is  smeared in a region around the insertion points; since we
are concerned with the vicinity of $x_1$ we just smear the source at this
point and do not concern ourselves with the other sources. Let the smeared
field have a profile $f(x)$, with
\be
\int f(x) dx =1\, .
\ee
It is important that as we shrink the size of the smearing region the result 
of the integral should cease to depend on the size and shape of the smearing
region, and should be sensitive only to the total integral of $f$ given in
the above equation.

(2) \quad There are  four derivatives in the term of interest. The
directions of differentiation are summed indices, and so these derivatives
can be in the $z_0$ direction or in the $z_i$ directions. The strongest
potential divergences arise from the case where the derivatives are in the
$z_0$ direction. The reason is that the term $\phi_2$ is given by
\be
\phi_2(z_0,x)=K(z_0,x;x_2)
\ee
and $K$ is constructed to be a Dirichlet boundary to bulk propagator. Thus 
for $x\ne x_2$
\be
\lim_{z_0\rightarrow\epsilon} K(z_0,x,x_2)=0
\ee
and a more careful analysis will give
\be
\phi_2(z_0,x)\sim\partial_{x_i}\phi_2(z_0,x)\sim (z_0-\epsilon)
\ee
for $z_0\rightarrow\epsilon$. The above fact improves the convergence of the
terms that involve derivatives in the directions $z_i$, and so we will
concentrate below on the case where all derivatives are in the direction
$z_0$.

(3) \quad The term $t^{k_1}$ in eq. (3.29) is
\be
t^{k_1}(z_0,x)=\int dy\ K(z_0,x;y)f(y) 
\ee
where we will assume that since we are looking at short distances $K$ has 
the form of the flat space kernel given above. $x,y$ are 4-dimensional
variables, and $f$ is the smearing function introduced above.
For $z_0\rightarrow \epsilon$, we may expand
\bea
\int dy\  K_f(z_0, x; y) f(y)  ~~~~~~~~~~~~~~~~~~~~~~~~~~~~~~~~~~~~\nonumber \\
= \int dy\ 
K_f(z_0,x;y)[f(x)+(y-x)_if_{,i}(x) +(y-x)_i(y-x)_jf_{,ij}(x)+\dots]
\eea
This equals
\be
\alpha f(x)+\beta (z_0-\epsilon)^2 f_{,i}{}^i(x)+\dots
\ee
where $\alpha, \beta$ are of order unity, and the term of order 
$z_0-\epsilon$ vanishes by the rotational symmetry of $K_f$. (This
vanishing will be important in what follows.)

Thus when we compute $z_0$ derivatives of $\int K f $ we get
\be
\partial_{z_0}\int K_f(z_0,x;y)f(y)dy=(z_0-\epsilon)f_{,i}{}^i(x)\rightarrow
0 ~~{\rm for}~~z_0\rightarrow \epsilon
\ee

(4) \quad 
Let us define
\be
h(z_0,x)\equiv
\partial_{z_0}\phi^{k_2}(z_0,x)\partial_{z_0}\phi^{k_3}(z_0,x)
\ee 
as mentioned above, $h$ is a smooth bounded function in the region near $x_1$
that we are studying.  Now  the integral we are studying has the form
$$\int h(z_0,x)\partial_{z_0}^2 K(z_0,x;y) f(y)dz_0 dx dy$$
and we are considering the integral in a box in 5-d space, enclosing the 
point $x_1$ of the boundary but not enclosing $x_2, x_3$.
Let us integrate once by parts. Then we get
\be
\int_\partial  h(\epsilon,x)\lim_{z_0\rightarrow\epsilon}\partial_{z_0}
K_f(z_0,x;y) f(y) dx dy-\int \partial_{z_0}
h(z_0,x)\partial_{z_0}K_f(z_0,x;y) f(y)dz_0 dx dy
\ee 
The first term vanishes by what we found above. The second term we integrate
by parts again to get
\be
-\int_{\partial} \partial_{z_0} h(z_0,x)\lim_{z_0\rightarrow
\epsilon}K_f(z_0,x;y) f(y) dx dy+\int \partial_{z_0}^2 h(z_0,x)K_f(z_0,x;y)
f(y)dz_0 dx dy
\ee 
This is
\be
-\int_{\partial} \partial_{z_0} h(z_0,x)f(x) dx +\int \partial_{z_0}^2
h(z_0,x)K_f(z_0,x;y) f(y)dz_0 dx dy
\ee 
Note that the smeared function must satisfy $\int f(x) dx =1$.
Since $h$ is smooth, we may approximate the first term in the above
equation to get
\be
-\partial_{z_0} h(z_0,x_1) +\int \partial_{z_0}^2 h(z_0,x)K_f(z_0,x;y)
f(y)dz_0 dx dy
\ee 
The bulk integral in the above is now no more singular than the
two-derivative term in the initial action(3.29).

Note that we have not used any moments of the smearing function apart from 
its zeroth moment; this is important because as mentioned above if there
was any additional dependence on $f$ then the functional derivative was ill
defined, and the result would have to be called regulation dependent.

(5) \quad By what was said at the start of this Appendix, we thus
understand how to regulate the boundary terms that arise in any
integrations by parts in the calculation of the extremal correlator.
Instead of explicitly smearing the source function, we can imagine using a
momentum basis to expand functions in a region around $x_1$: This region
would have to be much larger than the AdS length scale $R$ for the Fourier
transform to make sense, but can be taken to be much smaller than the
distance to the other insertions at $x_2, x_3$. This enables a simple
calculation, while avoiding any contact terms that can arise when $x_1$
approaches $x_2, x_3$. The result of this calculation was presented in
section 3.

\section*{Acknowledgments}
It is a pleasure to aknowledge useful conversations and correspondence with
Massimo Bianchi, Elena Caceres, Marc Grisaru, Juan Maldacena, Robert McNees,
Oliver DeWolfe,  Mukund Rangamani, Witold Skiba, and Leonard Susskind.
E.D'H gratefully acknowledges the hospitality of the Institute for Theoretical
Physics in Santa Barbara and of the Aspen Center for Physics, where part of this
work was carried out. 

The research of E.D'H is supported in part by NSF Grants No.
PHY-95-31023 and PHY-98-19686, D.Z.F.  by NSF Grant No. PHY-97-22072,  
S.D.M., A.M. and L.R. by D.O.E. cooperative 
agreement DE-FC02-94ER40818. L.R. is supported in part 
by INFN `Bruno Rossi' Fellowship.


\begin{thebibliography}{ll}

\bibitem{maldacena}J. Maldacena, `The Large $N$ Limit of Superconformal
Theories and Supergravity', Adv.Theor.Math.Phys. {\bf 2} (1998) 231-252, 
hep--th/9711200.

\bibitem{polyakov}{S.S. Gubser, I.R. Klebanov and A.M. Polyakov,
`Gauge Theory Correlators from Non--critical String Theory', Phys.Lett. 
{\bf B428}  (1998) 105-114,
hep--th/9802109.}

\bibitem{witten}{E. Witten, `Anti--de Sitter
Space and Holography', Adv.Theor.Math.Phys. {\bf 2} (1998) 253-291,  
hep--th/9802150.}

\bibitem{review}
O.~Aharony, S.S.~Gubser, J.~Maldacena, H.~Ooguri and Y.~Oz,
``Large N field theories, string theory and gravity,''
hep-th/9905111.

\bibitem{seiberg}
S.~Lee, S.~Minwalla, M.~Rangamani and N.~Seiberg,
``Three point functions of chiral operators in D = 4, N=4 SYM at large N,''
Adv. Theor. Math. Phys. {\bf 2}, 697 (1998)
hep-th/9806074.

\bibitem{april}D.Z.~Freedman, S.D.~Mathur, A.~Matusis and L.~Rastelli,
``Correlation functions in the CFT(d) / AdS(d+1) correspondence,''
Nucl. Phys. {\bf B546}, 96 (1999)
hep-th/9804058.

\bibitem{texas}
R.~Corrado, B.~Florea and R.~McNees,
``Correlation functions of operators and Wilson surfaces in the d = 6, (0,2)
                  theory in the large N limit,''
hep-th/9902153.

\bibitem{liutseytlin3}H.~Liu and A.A.~Tseytlin,
``Dilaton - fixed scalar correlators and AdS(5) x S**5 - SYM
                  correspondence,''
hep-th/9906151.

\bibitem{ferrara}
L.~Andrianopoli and S.~Ferrara,
``On short and long 
SU(2,2/4) multiplets in the AdS / CFT correspondence,''
hep-th/9812067.

\bibitem{bianchi}
M.~Bianchi, S.~Kovacs, G.~Rossi and Y.S.~Stanev,
``On the logarithmic behavior in N=4 SYM theory,''
hep-th/9906188.

\bibitem{skibanew}
W.~Skiba,
``Correlators of short multitrace operators in N=4 supersymmetric Yang-
                  Mills,''
hep-th/9907088.

\bibitem{trivedi}
P.~Kraus, F.~Larsen and S.P.~Trivedi,
``The Coulomb branch of gauge theory from rotating branes,''
JHEP {\bf 03}, 003 (1999)
hep-th/9811120.

\bibitem{klebwit}
I.R.~Klebanov and E.~Witten,
``AdS / CFT correspondence and symmetry breaking,''
hep-th/9905104.


\bibitem{dobrev}
V.K.~Dobrev and V.B.~Petkova,
``All Positive Energy Unitary Irreducible Representations Of Extended
                  Conformal Supersymmetry,''
Phys. Lett. {\bf 162B}, 127 (1985).


\bibitem{vannieuv}{H. J. Kim, L. J. Romans, and P. van Nieuwenhuizen,
`The Mass Spectrum Of Chiral $N=2$ $D=10$ Supergravity on $S^5$',
Phys. Rev. {\bf D32} (1985) 389.}


\bibitem{dhfskiba}
E.~D'Hoker, D.Z.~Freedman and W.~Skiba,
``Field theory tests for correlators in the AdS / CFT correspondence,''
Phys. Rev. {\bf D59}, 045008 (1999)
hep-th/9807098.

\bibitem{int}
K.~Intriligator,
``Bonus symmetries of N=4 superYang-Mills correlation functions via AdS
                  duality,''
Nucl. Phys. {\bf B551}, 575 (1999)
hep-th/9811047.


\bibitem{intskiba}
K.~Intriligator and W.~Skiba,
``Bonus symmetry and the operator product expansion of N=4 SuperYang-Mills,''
hep-th/9905020.

\bibitem{howwest}
B.~Eden, P.S.~Howe and P.C.~West,
``Nilpotent invariants in N=4 SYM,''
hep-th/9905085.

\bibitem{petkou}
A.~Petkou and K.~Skenderis,
``A Nonrenormalization theorem for conformal anomalies,''
hep-th/9906030.


\bibitem{russians}
G.~Arutyunov and S.~Frolov,
``Some cubic couplings in type IIB supergravity on AdS(5) x S**5 and three
                  point functions in SYM(4) at large N,''
hep-th/9907085.

\bibitem{lee}
S.~Lee,
``AdS(5) / CFT(4) four point functions of chiral primary operators: Cubic
                  vertices,''
hep-th/9907108.

\bibitem{gunmarc}
M.~Gunaydin and N.~Marcus,
``The Spectrum Of The $S^5$ Compactification Of The Chiral N=2, D = 10
                  Supergravity And The Unitary Supermultiplets Of U(2,
                  2/4),''
Class. Quant. Grav. {\bf 2}, L11 (1985).

\bibitem{scalar}
E.~D'Hoker and D.Z.~Freedman,
``General scalar exchange in AdS(d+1),''
Nucl. Phys. {\bf B550}, 261 (1999)
hep-th/9811257.



\end{thebibliography}
\end{document}